\documentclass[10pt, conference, compsocconf]{IEEEtran}
  \usepackage{cite}
 \usepackage{booktabs}

\usepackage{graphicx}
 \usepackage{epstopdf}
\usepackage[caption = false]{subfig}

\usepackage{multicol}
\usepackage{multirow}
\usepackage{array}

\usepackage{algorithm,algorithmic}
\newcommand{\SWITCH}[1]{\STATE \textbf{switch} (#1)}
\newcommand{\ENDSWITCH}{\STATE \textbf{end switch}}
\newcommand{\CASE}[1]{\STATE \textbf{case} #1\textbf{:} \begin{ALC@g}}
\newcommand{\ENDCASE}{\STATE \textbf{end case} \end{ALC@g}}
\newcommand{\INPUT}{\STATE \textbf{Input:}}
\newcommand{\PROCEDURE}[2]{\STATE \textbf{procedure} #1(#2) \begin{ALC@g}}
\newcommand{\ENDPRO}{\STATE \textbf{end procedure} \end{ALC@g}}
\newcommand{\FUNCTION}[2]{\STATE \textbf{function} #1(#2) \begin{ALC@g}}
\newcommand{\ENDFUNC}{\STATE \textbf{end function} \end{ALC@g}}

\usepackage{xcolor}
\usepackage{listings}
\usepackage{fancyvrb}
\definecolor{dkgreen}{rgb}{0,0.6,0}
\definecolor{gray}{rgb}{0.5,0.5,0.5}
\definecolor{mauve}{rgb}{0.58,0,0.82}

\begin{document}
\newcommand{\head}[1]{\textnormal{\textbf{#1}}}
%
\title{Asynchronous progress design for a MPI-based PGAS one-sided communication system}


\author{\IEEEauthorblockN{Huan Zhou}
\IEEEauthorblockA{High Performance Computing Center Stuttgart (HLRS)\\
University of Stuttgart\\
Germany\\
zhou@hlrs.de}
\and
\IEEEauthorblockN{Jos\'e Gracia}
\IEEEauthorblockA{High Performance Computing Center Stuttgart (HLRS)\\
University of Stuttgart\\
Germany\\
gracia@hlrs.de}
}


%


\maketitle

\begin{abstract}
Remote-memory-access models, also known as one-sided
communication models, are becoming an interesting alternative to
traditional two-sided communication models in the field of High
Performance Computing. 
In this paper we extend previous work on an MPI-based, locality-aware
remote-memory-access model with a asynchronous progress-engine for
non-blocking communication operations. Most previous related work
suggests to drive progression on communication through an additional
thread within the application process. In contrast, our scheme uses an
arbitrary number of dedicated processes to drive asynchronous
progression. 
Further, we describe a prototypical
library implementation of our concepts, namely DART, which is used to
quantitatively evaluate our design against a MPI-3 baseline reference.
The evaluation consists of micro-benchmark to measure overlap of
communication and computation and a scientific application kernel to
assess total performance impact on realistic use-cases. Our benchmarks
shows, that our asynchronous progression scheme can overlap 
computation and communication efficiently and lead to substantially
shorter communication cost in real applications.
\end{abstract}

\begin{IEEEkeywords}
DART; MPI; one-sided; asynchronous progress; data-locality; overlap

\end{IEEEkeywords}

%
\IEEEpeerreviewmaketitle

\section{Introduction}
Towards asynchronous communication operations, MPI standard~\cite{mpi3-standard} 
defines an imprecise
progress rule for implementors~\cite{journals/ijhpca/BrightwellRU05}.
However, different interpretations (\textit{strict} or \textit{weak}) 
of progress rule could lead to
differing progress patterns of the non-blocking communication operations
(include RMA communication routines).

\begin{figure*}
\begin{center}
\subfloat[\textit{Strict} interpretation]{\includegraphics[width=0.43\textwidth,height=0.21\textheight]{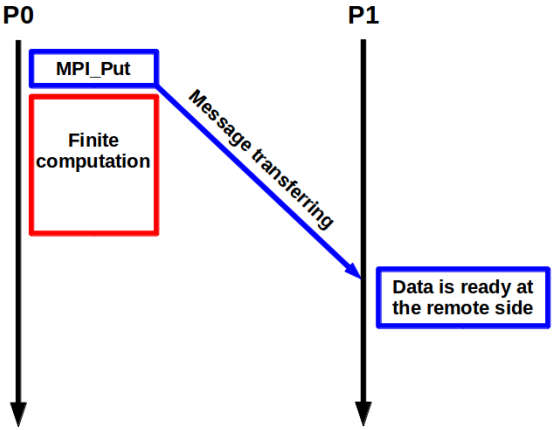}\label{strict}} 
\subfloat[\textit{Weak} interpretation]{\includegraphics[width=0.43\textwidth,height=0.21\textheight]{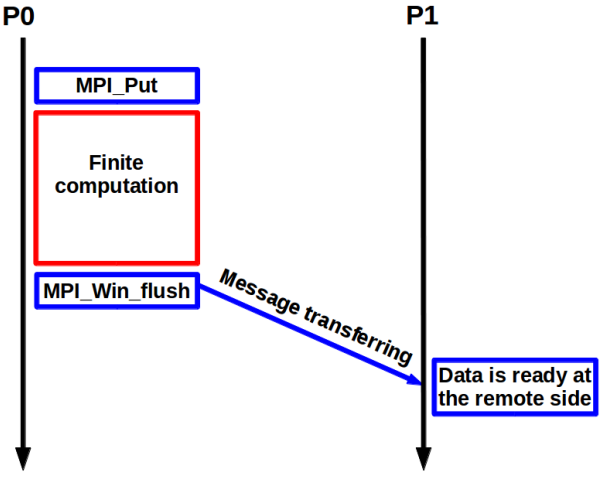}\label{weak}}
\end{center} 
\caption{Progress patterns of {\em MPI\_Put}
in terms of \textit{strict} and \textit{weak} interpretation.}
\label{progress}
\end{figure*}

Figure~\ref{progress}, take {\em MPI\_Put} for example,
describes the view of 
how MPI RMA communication operations implement
according to the 
\textit{strict} and \textit{weak} interpretation, respectively.
As Fig.~\ref{progress}\protect\subref{strict}
shows, once the non-blocking put operation
has been posted on the origin process (P0),
the data transfer can be enabled
independent of the further MPI synchronization calls
(e.g., \textit{flush}es) at the P0 side.
This pattern supports the truly asynchronous 
completion of communications
(i.e., asynchronous progression)
by offering the overlap of communication and computation.
On contrary, in Fig.~\ref{progress}\protect\subref{weak},
the put communication is delayed until
the ensuring \textit{flush} call happens at the origin.
This interpretation depends on MPI synchronization call from P0
to make explicit progress.
{\em MPI\_Get} and the request-based
RMA ({\em MPI\_Rput/Rget}) operations exhibit the similar behavior 
difference as {\em MPI\_Put} does
under the above two interpretations. 
The MPI programmers should be aware that MPI RMA
communications do not overlap with computation with certainty
in all MPI implementations. Clearly,
allowing overlap of communication and computation in parallel
applications, especially for long messages,
is beneficial.
This helps to
reduce host processor overhead by making the host processor
less involved in the transmission or reception of data
and to hide latency by letting CPU contribute to the computation
in the interim.
Therefore, an implementation adheres to the 
\textit{strict} interpretation has performance advantage 
potential over one  that supports the \textit{weak} 
interpretation for the applications 
with some degree of overlap potential~\cite{journals/corr/abs-1302-4280}. 

Traditionally, two methods -- multi-threading-based (refer to MPICH, MVAPICH and Intel MPI) and 
kernel thread-based (refer to Cray MPI~\cite{Cray-asynchronous}) -- are well-studied by 
scientists to support for
communication/computation overlap to 
minimize the host overhead~\cite{conf/ipps/SiPHBTI15}.
However, the disadvantages inherent in them gradually 
become impediment to scalable application performance.
Additionally, an innovative approach -- process-based~\cite{PGAS-MPI/Designalternatives,
conf/ipps/SiPHBTI15}, is devised to designate arbitrary number of cores
per node to be the processes handling the communication progression.
It has been proved to be an efficient alternative
method supporting asynchronous communications
for current multi-/many-core architecture,
compared with the conventional methods.

DART~\cite{dash, conf/pgas/ZhouMIGGF14} -- as an MPI3-based PGAS runtime system
-- is designed to provide one-sided communication operations 
(including blocking and nonblocking) with data locality-awareness
in mind.
It can directly be adopted by users or easily forms the basis
for the runtime time system of certain PGAS project.
In addition, compared to MPI codes,
DART codes are more concise and easier-to-read
due to that DART internally hides the MPI complexities from users~\cite{conf/LHAM/zhou}.
Given the performance properties,
PGAS languages require of the non-blocking RMA interfaces
to hide network latencies by overlapping
communication and computation.
Therefore, supporting the asynchronous progression
in DART RMA is inevitable.

In section~\ref{sec:desgin} we illustrate the design method
of DART asynchronous progression engine with the process-based
approach and then theoretically analyze the efficiency
and suitability of applying the DART non-blocking
RMA communication operations on parallel applications.
In section~\ref{sec:evaluation} we evaluate the 
performance of DART non-blocking RMA communication operations
with micro-benchmarks and 3D heat conduction application as a 
representation for a load of numerical simulation schemes.
We compare the DART RMA and the native MPI RMA in all experiments.


\section{Designing the DART progress engine}
\label{sec:desgin}
In this section, we outline the design details
which address the process-based implementation issues
regarding the asynchronous
progression engine in DART. Foremost, we partition cores on a node
between progress processes and user application processes.
Next, we introduce the way of setting up global memory (similar to MPI window)
across the partitioned cores.
We then internally launch the asynchronous progression engine
by intercepting and overriding the DART RMA functions.
Finally, the suitability of the appliance of DART non-blocking
RMA communication operations on parallel applications is discussed.

\subsection{Core partitioning}

On the premise that the DART global memory architecture is applied~\cite{conf/europar/ZhouIG15}, 
processes within the same shared memory domain are possible to
communicate via direct load/store accesses.
Detailedly, the user
memory regions are directly
mapped into the on-node progress processes' memory address spaces.
This property of DART system positively affects the appliance 
of the process-based approach
in a way of 
reducing the time complexity. 
Consequently, 
with the mapped memory address, RMA operations can be 
redirected and handled by the progress processes 
without the verbose message transfers between 
the application (user) processes and the progress processes within
the same shared-memory domain.

\begin{figure}[H]
\begin{center}
\includegraphics[width=0.85\linewidth]{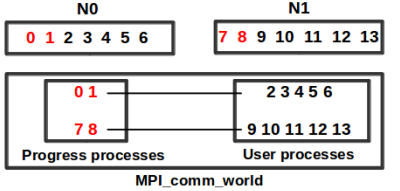}
\end{center} 
\caption{DART process layout with asynchronous progression enabled.
The two progress processes are marked with red ink
and serve the RMA requests originating from other on-node 
application processes. A node signifies a shared memory domain.}
\label{progresslayout}
\end{figure}

At DART initialization time (i.e., in {\em dart\_init}),
one launches applications
with a number of processes. 
Several processes within each node
are then designated to be the progress processes for
the other processes located in the same node.
Figure~\ref{progresslayout} shows a specific instance
where we use two progress processes within each node
and each of them is pinned to a core.
Additionally, the progress processes are
designated to on-node application
processes at DART initialization time
as evenly as possible.
Actually this asynchronous progression engine
allows one to generate arbitrary number of progress 
processes internally on-demand.
Importantly, after excluding the progress processes, 
the remaining processes comprise the
{\em DART\_TEAM\_ALL} that are only visible to the user applications.
Apparently, 
the progress processes are only visible to the DART
system instead of the users.
Consequently, the progress processes are prohibited 
to be explicitly involved in all DART interfaces except the 
{\em dart\_init} or {\em dart\_exit} (DART termination routine).
Specifically, inside the {\em dart\_exit},
the progress processes are busy waiting for the 
commands delivered by other on-node user processes in an 
{\em MPI\_Iprobe} loop.
The asynchronous progression engine
ensures that the progress processes react properly according to
the command characteristics, which
will be discussed in the subsections below.
\subsection{Global memory setup}
The progress processes are natively designed 
to be invisible to the target applications.
However, we here let progress processes being
aware of any global memory allocation
since the progress processes, after all, would be 
a proxy for the application processes to
asynchronously perform the RMAs to 
the target global memory segments.

The memory region reserved for non-collective global
memory allocation is statically built up at DART initialization time.
Meantime, the progress processes are intuitively involved
in the collective window creation operation.
On the other hand, we allocate the collective global memory region 
dynamically by collectively
invoking {\em dart\_team\_memalloc\_aligned} 
on certain team (similar to MPI communicator).
This interface is originally visible to application processes
instead of progress processes.
Therefore, one of the participated application processes internally
notifies all on-node progress processes.
After consuming the intra-node notifications,
the progress processes are involved in the collective global
memory allocation with the received parameter (i.e., \textit{index},
see Table~\ref{requestRMA}).
In this case, a notifying operation is equivalent to
a small intra-node message transferring.
It is simply performed by using pair-wise
{\em MPI\_Send} and {\em MPI\_Recv}, which
will not cost a lot.
Clearly, the progress processes are internally involved in all global 
memory allocations (window creation) within DART system.
To minimize the space complexity, 
I allocate
memory of 0 byte in the progress processes' memory space
as a result of memory mapping. 
This is feasible since the progress processes' memory space
would never be accessed by users. Therefore, there is no
extra memory allocation wasting from the application
point of view.
Importantly, the target global memory segment should be 
accessible to the progress processes after
the window (global memory segment) is generated.
Therefore, the progress processes must
immediately start a global lock epoch
to all processes related to this window
and end it when the window is destroyed.

\subsection{Asynchronous non-blocking RMA communication operations}
\label{non-blocking}
\begin{figure*}
\begin{center}
\includegraphics[width=0.85\textwidth, height=0.21\textheight]{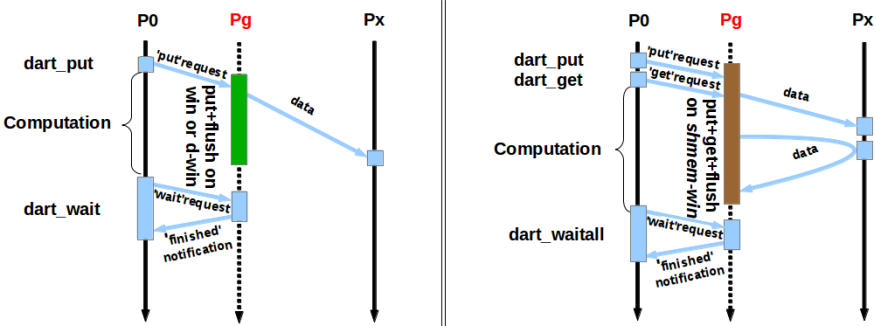}
\end{center} 
\caption{DART non-blocking intra-node and inter-node RMA communication protocols using process-based approach.
P0 is the origin which initiates the RMA operations, Px is the destination. Pg (marked with red ink)
denotes the progress process corresponding to P0. Left: P0 and Px are in the different nodes,
Pg consumes one RMA request. Right: P0 and Px are in same node, 
Pg consumes two contiguous RMA requests.}
\label{dart_progress}
\end{figure*}

DART non-blocking RMA operations can be activated successfully
as the processes are granted access to the window within
the above global lock epoch.
Figure~\ref{dart_progress} gives insights into the asynchronous progression design of
the DART non-blocking RMA communication operations.
In detail, first of all, the origin chooses an on-node progress process.
Instead of actually performing the 
RMA operations, it sends the RMA (i.e., put or get) requests to its
current progress process. 
This progress process then takes over the actual message transfers
meant for the given destination Px on MPI shared-memory window \textit{sharedmem-win}
or RMA window (i.e., MPI-created
window \textit{win} for non-collective global memory and 
MPI dynamically-created window \textit{d-win} for collective global memory)
according to whether the target process is an on-node application process or not.
In this sense, the origin can do its own computation tasks independently
from the initiated RMA communications.
It is clearly observed that the asynchronous progression engine
keeps the data locality in mind in order to achieve high performance.
The origin sends a 'wait' request 
to its progress process when invoking
{\em dart\_wait} or {\em dart\_waitall} and then busy-waits for the 
finished notification from its progress process.

Figure~\ref{dart_progress} shows two common scenarios
where the RMA requests are consumed by progress process
brokenly (shown in left) and consecutively (shown in right),
respectively.
The progress process would like to \textit{flush} the target immediately
if it finds no RMA request followed closely.
Otherwise, 
if the RMA requests are consecutively sent to
the progress process, ideally the progress process
will build a backlog of RMA requests.
They are processed in batch by progress process
when it receives a 'wait' request or becomes idle (i.e., probe nothing).
In this case, the data transfers fail to start as soon as
the RMA calls occur. However, the progression in the
accumulative RMA communications can be activated
automatically when the progress process probes no message,
which provides overlap of communication and computation to some extent.
Furthermore, this behavior can somehow bring performance improvement
through amortizing a \textit{flush} synchronization call with multiple
RMA operations. 

\begin{table}[!t]
\centering
\caption{Packet structure.}
\label{requestRMA}
\begin{tabular}{ll}
 \toprule[1.5pt]
\head{Field}                          & \head{Value} \\ 
 \midrule
\textit{dest}                           & The target process ID    \\			 
\textit{index}                 	 & Denote the involved team                                                                            \\ 
\textit{origin\_offset}                 & \begin{tabular}[c]{@{}l@{}}The offset relative to the beginning of \\ origin memory  segment\end{tabular} \\ 
\textit{target\_offset}                 & \begin{tabular}[c]{@{}l@{}}The displacement relative to the beginning \\ of target memory segment\end{tabular}  \\ 
\textit{data\_size}                     & The transferred data size                                                                                 \\ 
 \cmidrule(l){2-2}
\multirow{2}{*}{\textit{segid}}         & 0: Indicate the non-collective global memory                                                              \\ 
                               & $\geq$ 1: Indicate the specific collective global memory                                                         \\ 
                                \cmidrule(l){2-2}
\multirow{2}{*}{\textit{is\_shmem}} & 0: Origin and target are in different nodes                                                               \\ 
                               & 1: Origin and target are in the same node\\
                               
                                \bottomrule[1.5pt]
                               
\end{tabular}
\end{table}

The application processes, rather than the 
progress processes, are originally aware of
the raw information on the DART RMA communications.
Therefore, an application process should
encapsulate the raw information first
and then sends it as a packet to
its corresponding progress process.
Essentially,
the packet is determined in a way
of abstracting the input data
provided by DART RMA operations.
The input data basically consists of 
three attributes, that is, global pointer, message size,
and origin address, where the global pointer~\cite{conf/europar/ZhouIG15}
addresses the remote data item and thus
exposes the target unit/process ID, \textit{segid} (see Table~\ref{requestRMA}),
target address offset and \textit{index}.
The message size (\textit{data\_size}), \textit{segid} and \textit{index}
should be straightforwardly included into the packet.
Besides, the progress process should perform
RMA communications with locality-awareness in mind.
Therefore, a signal value (\textit{is\_shmem}) indicating that
data transfer occurs within node
or across nodes is entailed
in the packet.
After obtaining a local base pointer 
the raw origin address can be deduced by the progress process
with the aid of origin address offset (\textit{origin\_offset}).
Besides, the relative displacement (\textit{target\_offset}) that is passed
to the RMA operation should be included in the packet
for locating the target data.
Accordingly, the packet is represented
as the data structure shown in Table~\ref{requestRMA}.
It is identified as
a derived datatype through
a call to {\em MPI\_Type\_create\_struct}\cite{mpi3-standard}.

\begin{figure*}
\centering
 \begin{multicols}{2}
  \begin{algorithmic}[1]
   \scriptsize
   \INPUT global pointer p, message size m, target process dst
    \FUNCTION{dart\_get}{p, m, dst}
    \STATE g $\leftarrow$ TranslatetoPG;
    \STATE packet $\leftarrow$ EncodeComm(p, m, dst)
    \STATE MPI\_Send(packet, GET ...g)
    \STATE handle.g $\leftarrow$ g
    \STATE return handle
    \ENDFUNC
   \FUNCTION{dart\_wait}{handle}
    \STATE g $\leftarrow$ handle.g
    \STATE MPI\_Send(NULL, WAIT ...g)
    \STATE MPI\_Recv(NULL, WAIT ...g)
    \COMMENT{Indicates control message}
   \ENDFUNC
   \FUNCTION{dart\_waitall}{handle[], NUM}
    \STATE i = 0
    \WHILE{$(i++)<NUM$}
    \STATE g $\leftarrow$ handle[i].g
    \IF{g is targeted at first time}
     \STATE MPI\_Send(NULL, WAIT ...g)
     \STATE MPI\_Recv(NULL, WAIT ...g)
    \ENDIF
    \ENDWHILE
   \ENDFUNC
  \end{algorithmic}
 \end{multicols}
 \caption{Pseudo-code for application processes.}
 \label{applicationcode}
\end{figure*}

\begin{figure*}
 \centering
 \begin{multicols}{2}
 \begin{algorithmic}[1]
  \scriptsize
 \PROCEDURE{progress}{}
 \WHILE{$running$}
   \STATE $header \leftarrow Iprobe()$
  \IF{$header.flag$}
   \SWITCH{$header.messageType$}
          \CASE{$GET$}
          \STATE MPI\_Recv(packet, GET, ...header.o)
          \STATE buf $\leftarrow$ DecodeObuf(packet.origin\_offset)
          \STATE dst $\leftarrow$ packet.dest
          \STATE target\_offset $\leftarrow$ packet.target\_offset 
          \STATE win $\leftarrow$ DecodeWin(packet.is\_shmem, packet.segid, packet.index)
          \STATE MPI\_Get(buf, dst, target\_offset, ...win)
          \STATE request.dest $\leftarrow$ dst
          \STATE request.win $\leftarrow$ win
          \STATE q $\leftarrow$ AddQueue(request)
          \STATE break
          \ENDCASE
          \CASE{$WAIT$}
          \STATE MPI\_Recv(NULL, WAIT, ...header.o)
          \WHILE{q is not an empty queue}
          \STATE request $\leftarrow$ DeQueue
          \STATE MPI\_Flush(request.dest, request.win)
          \ENDWHILE
          \STATE MPI\_Send(NULL, WAIT, ...header.o)
          \ENDCASE
      \ENDSWITCH
 \ELSE
 \WHILE{q is not an empty queue}
 \STATE request $\leftarrow$ DeQueue
 \STATE MPI\_Flush(request.dest, request.win)
 \ENDWHILE
 \ENDIF
 \ENDWHILE
  \ENDPRO
 \end{algorithmic}
 \end{multicols}
 \caption{Pseudo-code for progress processes.}
 \label{progresscode}
\end{figure*}
 
Figures~\ref{applicationcode} and \ref{progresscode}
demonstrate the interactive activities between application processes
and the corresponding progress processes,
Fig.~\ref{applicationcode}
describes the algorithm for each of the DART get, wait
and waitall primitives straightforwardly.
Obviously, the progress process procedure shown
in Fig.~\ref{progresscode} is invoked as necessary
to make progress on outstanding RMA communications.
In this instance, get communication is implemented by
calling {\em MPI\_Get} and flush routine. 
The {\em MPI\_Rget}, coupled with {\em MPI\_Wait}
can actually be an efficient alternative.
For brevity we leave out the procedures
handling the collective memory allocation/deallocation, 
team creation/destroy operations, put operation (handled similar as get operation)
and termination routine (i.e., {\em dart\_exit}).

The eager protocol for small message transfers in MPI potentially
involves overlap of computation and communication (i.e.,
through buffering).
However, the rendezvous protocol for long
message transfers does not allow overlap between computation
and communication~\cite{smgprogress}.
To minimize the progression implementation overhead, especially for small messages,
we only enable asynchronous progress for DART non-blocking RMA communications
when the message sizes 
across the threshold value, which is determined in this paper
in terms of the data results below (refer to Sect.~\ref{host}).

\subsection{Suitability analysis}
It is significant to present the suitability
of this DART asynchronous progression approach
for the real-world applications. 
DART RMA communications are implemented based on 
the MPI passive-target RMA.
However, MPI implementations do not guarantee
truly passive RMA operations.
This means that the target can be involved
in an RMA operation only
by explicitly making MPI calls (any MPI call)
to ensure communication progression on its side.
Actually, the study~\cite{conf/ipps/SiPHBTI15} mentioned before 
adopted the process-based approach to realize the truly
passive RMA.
Besides, another study~\cite{conf/pvm/LiPHJP13} tries to implement
the truly passive RMA by using InfiniBand Atomics.
Achieving truly passive RMA operations is critical
for the irregular parallel algorithms.

Revising the Fig.~\ref{progress}\protect\subref{weak},
we can observe that
the RMA communications are essentially initiated
by letting the origin explicitly issue 
the synchronization calls even when 
the target process is ready.
In this case,
the actual data transfer will be deferred
and fails to overlap with the following computation
after employing the above two studies for truly
passive RMA. 
Detailedly, the performance of applications including
the traditional regular parallel algorithms
(e.g., stencil computation~\cite{conf/sc/CruzA14}, FFT~\cite{luszczek:06} and so on)
may benefit from this DART asynchronous progression
approach rather than the above two studies.
In addition, the regular parallel algorithms
feature balanced communication patterns and
still play an active role
in the engineering and scientific communities.

\section{Performance evaluation}
\label{sec:evaluation}
In this section, we present a comprehensive experimental 
evaluation and analysis of the asynchronous progress feature
discussed in the preceding sections.
The experimental environment was the Cray XC40 system at HLRS. 
It is
equipped with 7,712 compute nodes made up of dual twelvecore
Intel Haswell E5-2680v3 processor (one processor per
socket), which has exclusive 256 KB L2 unified cache for
each core. Therefore, each compute node has 24 cores running
at 2.5GHZ. The different compute nodes are
interconnected via a Cray Aries network using Dragonfly
topology. 

The first step in identifying the performance
of DART progress engine
is to evaluate the ability
to improve the application performance through
overlapping with computation for the non-blocking RMA 
communication operations. 
We should add that, 
the asynchronous progression feature is disabled
in Cray MPI by default and thus
should be turn on explicitly by setting
environment variables at compiler time.
(Surprisingly, after setting environment
variables as needed, it seems that the MPI asynchronous 
progression does work for the MPI non-blocking
point-to-point communications rather than
RMA communications.)
For a more comparative study,
a 3D heat conduction application benchmark
is then run and analyzed 
on DART and Cray MPI
implementations to investigate the synthetic impact 
of progression on the overlap of 
computation and communication.

In all experiments below we try to show the preliminary
advantages of DART non-blocking RMA operations in terms of 
hiding communication latencies with only two progress processes per node.
Furthermore, multiple trails with increasing number of progress processes
per node could be carried out to explore an optimal application performance.
Note that, the number of dedicated progress processes for 
computation-intensive applications
should be cautiously determined in order to
guarantee that there are a good number of computing
processes (application processes) to share the required workload.

\subsection{Host processor \textit{overhead} and application \textit{availability}}
\label{host}

\begin{figure*}
\begin{center}
\subfloat[MPI intra-node get]{\includegraphics[width=0.5\textwidth,height=0.2\textheight]{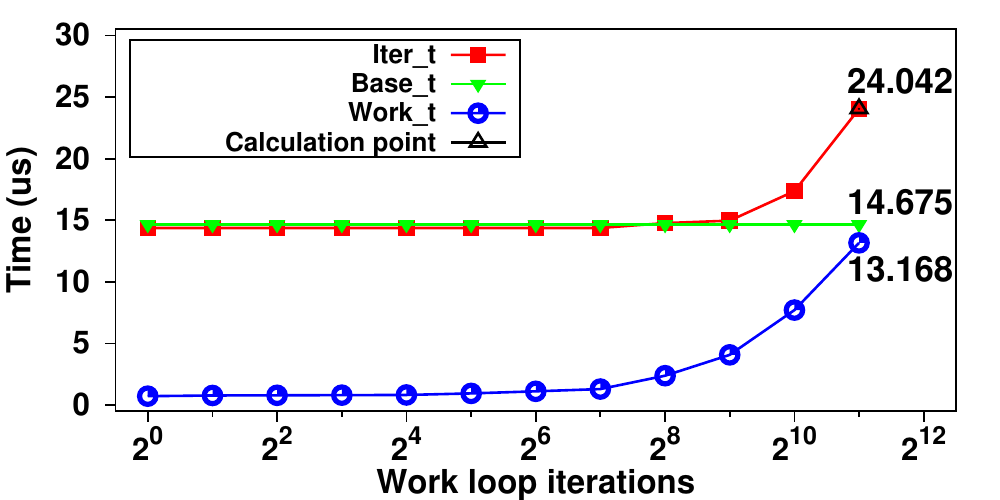}\label{SMBintrampiget}}
\subfloat[MPI inter-node get]{\includegraphics[width=0.5\textwidth,height=0.2\textheight]{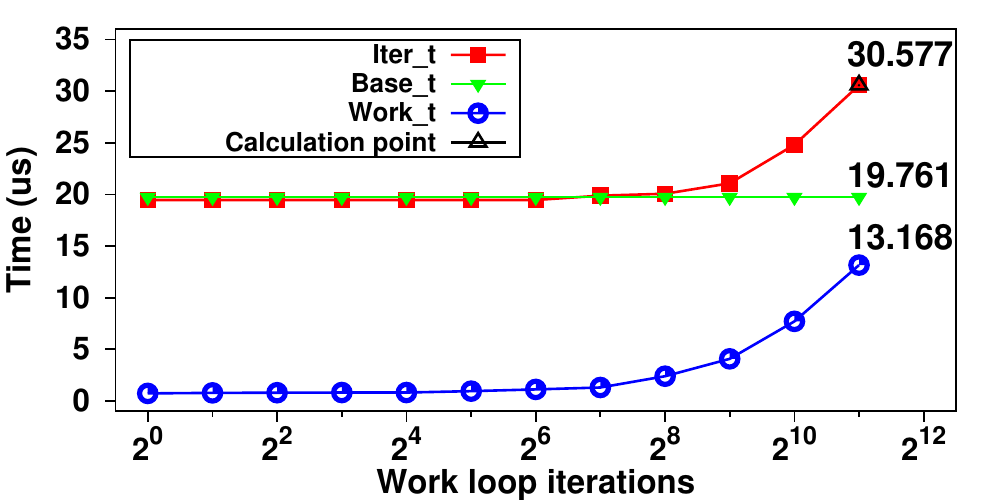}\label{SMBintermpiget}}
\end{center} 
\caption{The \textit{overhead} and application \textit{availability} achieved with MPI intra-node and inter-node get operations for 64 KB message sizes.
The measured \textit{overhead} is significant and the application \textit{availability} is $\sim$25.9\% for intra-node get and $\sim$11.9\% for inter-node get.
Without asynchronous progression.
}
\label{SMBmpiRMA}
\end{figure*}
\begin{figure*}
\begin{center}
\subfloat[DART non-blocking intra-node get]{\includegraphics[width=0.5\textwidth,height=0.2\textheight]{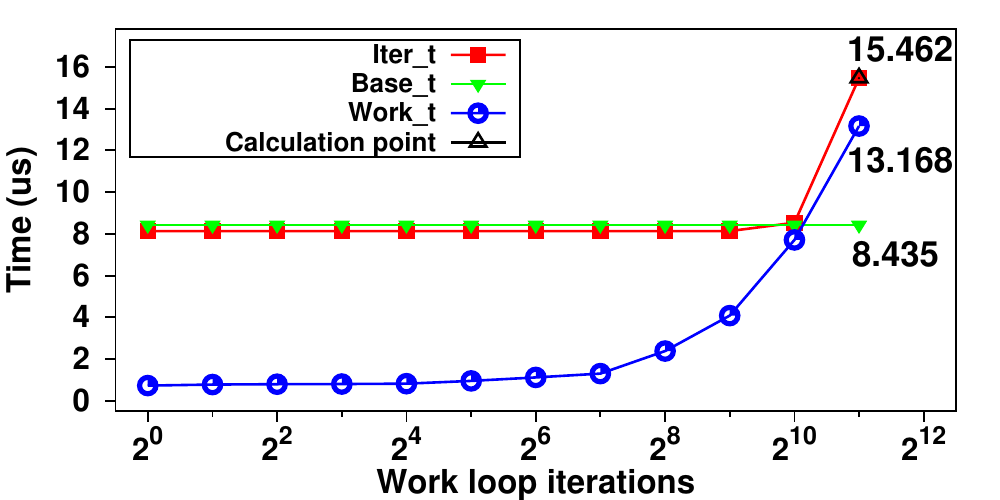}\label{SMBintradartget}}
\subfloat[DART non-blocking inter-node get]{\includegraphics[width=0.5\textwidth,height=0.2\textheight]{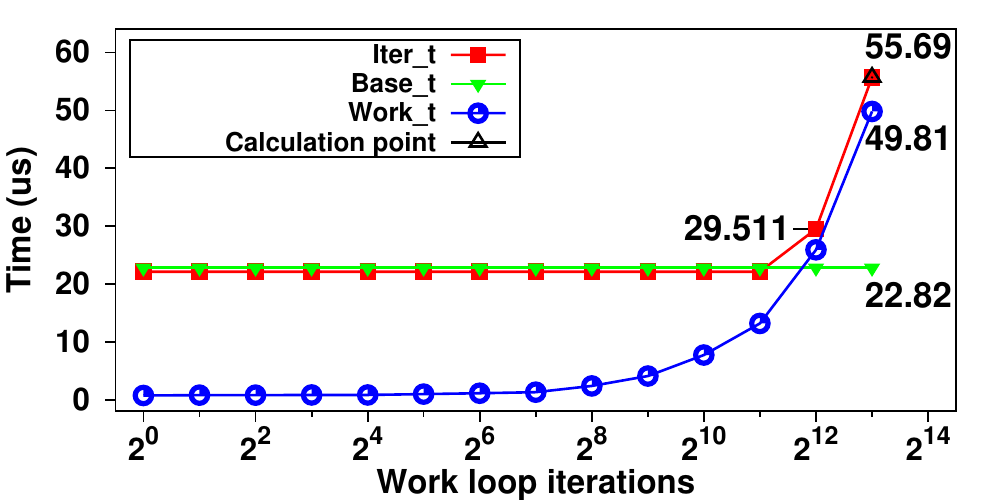}\label{SMBinterdartget}}
\end{center} 
\caption{The \textit{overhead} and application \textit{availability} achieved with DART non-blocking intra-node and inter-node get operations for 64 KB message sizes.
The measured \textit{overhead} is low and the application \textit{availability} is $\sim$72.8\% for intra-node get and $\sim$74.2\% for inter-node get.
With asynchronous progression.
}

\label{SMBdartRMA}
\end{figure*}
\begin{figure*}
\begin{center}
\subfloat[Non-blocking intra-node RMA]{\includegraphics[width=0.5\textwidth,height=0.26\textheight]{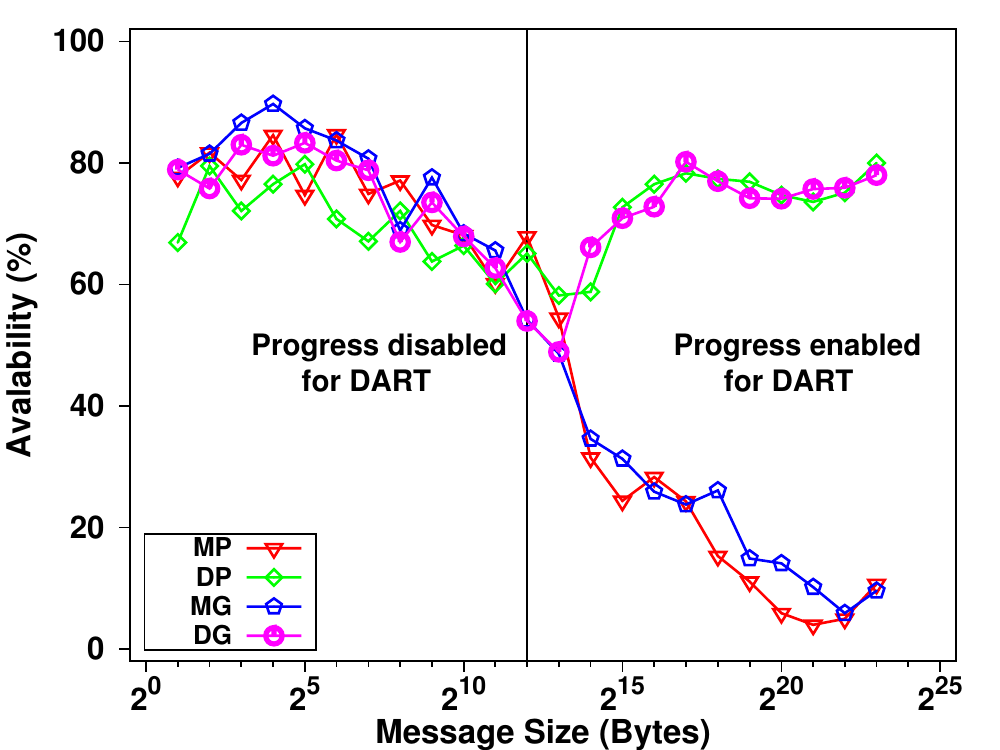}\label{intraavailability}}
\subfloat[Non-blocking inter-node RMA]{\includegraphics[width=0.5\textwidth,height=0.26\textheight]{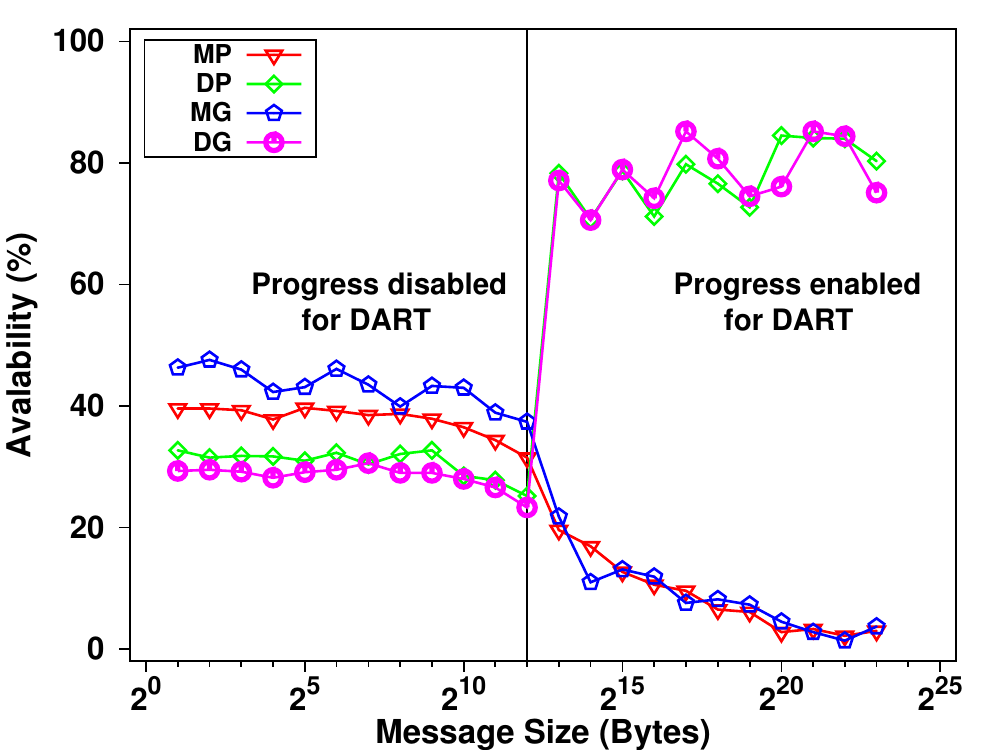}\label{interavailability}}
\end{center} 
\caption{A comparison of application \textit{availability} between DART non-blocking and MPI RMA operations.
"M" denotes MPI and "D" denotes DART. "P" denotes Put operation and "G" denotes Get operation.}
\label{availability}
\end{figure*}

In this section, we examine the overlap degree
of communication and computation when using
DART non-blocking RMA operations with 
asynchronous progression enabled,
based on the measurements modified from Sandia MPI
Micro-Benchmark Suite (SMB)~\cite{sandia}. Meantime,
we check the potential of DART non-blocking 
RMA communication operations to improve the application performance 
through overlapping with computation, 
compared with the Cray-MPI RMA (without asynchronous progression).

Specifically, we use a test component which is contained in the 
SMB. In this component, two basic metrics -- the host processor
\textit{overhead} and application \textit{availability}~\cite{SMBgroupmeeting} are 
measured for non-blocking MPI send and receive operations.
To target the non-blocking RMA operations,
we adjust this measurement by replacing the send or receive
routines with the non-blocking RMA routines.
The \textit{overhead} is defined as the amount
of time that a process is engaged in the transfer
of each message.
On the other hand, application \textit{availability}
is defined to be the fraction of total transfer time that
the application is free to perform non-MPI related work~\cite{sandia}.
The test measures the time to a loop where each iteration performs a non-blocking RMA
operation of a given size, defined workload and then
waits for the message transfer to complete. 
Detailedly, {\em MPI\_Get}
and {\em MPI\_Win\_flush} are used in MPI version
while {\em dart\_get} (the progress process performs {\em MPI\_Get} inside)
and {\em dart\_wait} are invoked in DART version.
The total amount of time (\textit{iter\_t}), the work time (\textit{work\_t})
and message transfer time (\textit{base\_t}) are assessed averagely in the test.
The test is repeated for increasing number of work loop iterations
and stops when {\textit{iter\_t}} hits more than 1.5 times {\textit{base\_t}}.
Note that, the \textit{overhead} and application \textit{availability}
are calculated at the stop point with the method of 
$overhead=iter\_t-work\_t$ and $availability=1-overhead/base\_t$.

Taking message size of 64 KB and get operation
for example, Figure~\ref{SMBmpiRMA}
illustrates the \textit{overhead} and application \textit{availability}
when using MPI get communication operations during one trial,
in intra-node and inter-node case.
The \textit{overhead} and application \textit{availability}
implied in these figures are baseline values
without featuring the asynchronous progression.
Likewise, the \textit{overhead} and application \textit{availability}
achieved with DART non-blocking get operations in one trial at message size of 64 KB
are shown in Figure~\ref{SMBdartRMA}.
We can clearly see that
DART non-blocking get communication operations
support smaller \textit{overhead}
and higher application \textit{availability} in all cases,
compared to the baseline values.
Obviously, at the work loop iterations
of 2048,
DART non-blocking get operations deliver less \textit{iter\_t}
than MPI get operations in the case of 
intra-node by partly hiding the data transfer time.
As we expected, DART-MPI get operations show less \textit{base\_t} than MPI
get operations due to the usage of \textit{shmem-win}, which also plays a part in 
improving the \textit{iter\_t}. For inter-node data transfers
DART non-blocking get operations can even show less \textit{iter\_t}
at 4096 than MPI get operations at 2048.

Figure~\ref{availability} comparatively
reports the application \textit{availability} achieved with the 
DART non-blocking and MPI RMA operations as a function of 
the message size by repeating the above measurement.
DART non-blocking RMA operations support slightly lower 
application \textit{availability} when the transferred message
sizes are not larger than 4 KB, which is in our expectation
since the asynchronous progression feature is disabled in this case.
Clearly, 
DART non-blocking RMA can get consistently better application
\textit{availability} than MPI RMA once the message sizes 
exceed 4 KB.
MPI RMA operations show very poor application
\textit{availability} 
for message sizes larger than 1 MB.
In addition,
a sudden drop occurs at the message size of 8 KB
in MPI RMA \textit{availability}.
This is also the reason for our enabling of 
DART asynchronous progress when message sizes are 
beyond 4 KB.
\subsection{3D heat conduction}
\begin{figure}[H]
\begin{center}
\includegraphics[width=\linewidth,height=0.26\textheight]{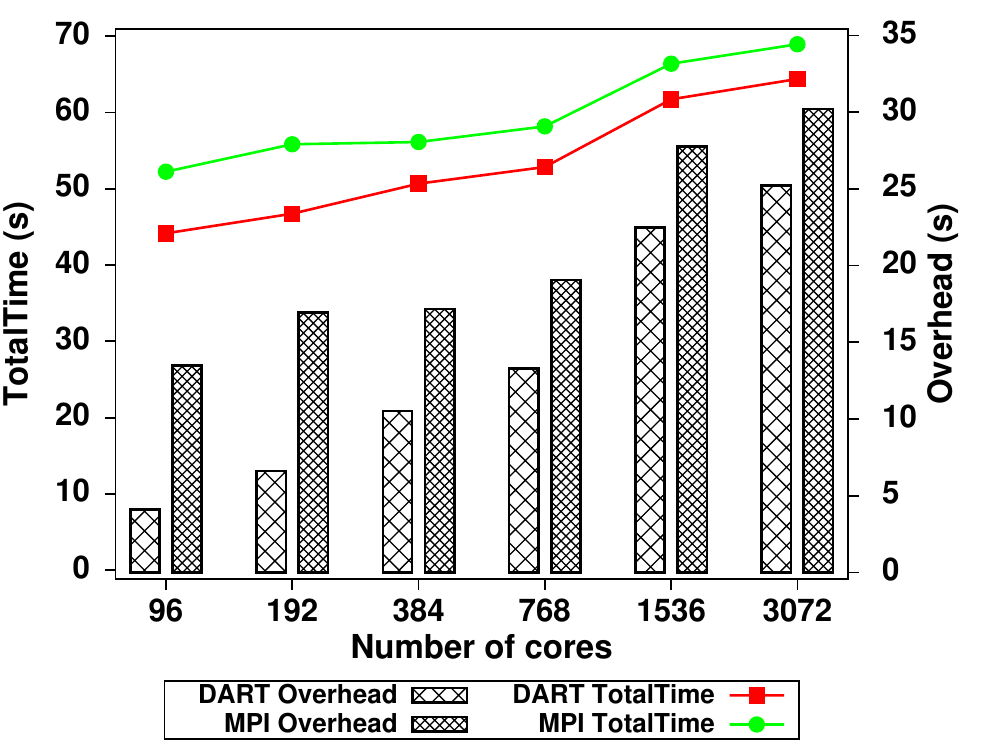}
\end{center} 
\caption{3D heat conduction performance comparison.}
\label{heatconduction}
\end{figure}
Heat conduction plays an important role in engineering communities.
It is a mode of heat transfer owning to molecular activity.
We simulate the phenomenon of 3D heat conduction in solids
with temperature-dependent thermal diffusivity based on an application
code parallelized with MPI point-to-point~\cite{3Dheat}. 
In this benchmark,
parallelization is done based on the checkerboard domain decomposition.
For our evaluation, we port this implementation to DART 
one-sided directives. Also, we implement this 3D heat conduction algorithm
based on MPI one-sided interfaces using passive target mode as the memory
synchronization mechanism for a fair comparison.
The boundary exchange is achieved by using non-blocking get operations.
Particularly, {\em MPI\_Rget} and {\em MPI\_Waitall} are
invoked in MPI implementation
while {\em dart\_get} (the progress process
performs {\em MPI\_Rget} inside) and {\em dart\_waitall} are used
in DART implementation. We stop the calculation
after 5000 iterations and plot the average data results of 15 runs 
with small execution time variation reported.

A weak scaling calculation is presented in the Figure~\ref{heatconduction}
with grid sizes varying from (132$\times$128$\times$2048)
to (132$\times$4096$\times$2048).
In this experiment, not only do we provide
the total time performance, but we also test 
the message transmission time (also denoted as
\textit{overhead} in Fig.~\ref{heatconduction}) that the CPUs are engaged in.
The DART implementation provides an average speedup of 1.122x
on processes ranging from 96 to 3072.
Although DART implementation dedicates less computing cores compared to
the MPI implementation, 
it is still clearly observed that better total time performance can be 
achieved by using DART implementation due to that the majority of network 
latency has been hidden. Likewise,
DART implementation makes good use of CPU for useful calculation and thus
achieves less \textit{overhead} via overlapping computation and communication.
On average, the CPUs in MPI implementation spends 39\% more time transmitting message
than in DART implementation.
In comparison to MPI implementation with the 
calculation constitutes an average of 65.8\% of total time,
the calculation in DART implementation takes an average of 75.8\% of the 
total time.

\section{Conclusion}
This paper has illustrated the design
of enabling the asynchronous communications of
DART non-blocking RMA for large message sizes by using the progress
process-based approach.
Such asynchronous progression can provide better
communication and computation degree of overlap
and meantime is designed to take the data locality
into consideration.
We have given a detailed evaluation
on the performance improvement
offered by the design using a Sandia MPI Micro-Benchmark (SMB)
and a 3D heat conduction application benchmark.
The results demonstrate that the communication-computation overlap 
can be efficiently achieved for large message sizes.
According to the SMB evaluation for host processor \textit{overhead},
the DART non-blocking RMA provides
lower host processor \textit{overhead}
and higher \textit{availability}
than the MPI RMA does.
Furthermore,
the 3D heat conduction application evaluation shows that
using DART asynchronous non-blocking RMA operations 
produces less calculation time than
using MPI RMA interfaces for up
to 3072 processes by partly hiding the communication latencies.



\section*{Acknowledgment}
This work has been supported by the European Community through the
project Polca (FP7 programme under grant agreement number 610686) and
Mont Blanc 3 (H2020 programme under grant agreement number 671697). We
gratefully acknowledge funding by the German Research Foundation (DFG)
through the German Priority Programme 1648 Software for Exascale
Computing (SPPEXA) and the project DASH.


%


\bibliographystyle{IEEEtran}
\bibliography{IEEEabrv,./paper}

\end{document}